\documentclass[12pt,preprint]{aastex}
\usepackage{graphics,graphicx,rotating,amsmath}

\newcommand{\lr}[1]{\left(#1 \right)}
\newcommand{\lrt}[1]{\left< #1 \right>}
\begin{document} 
\title{Consequences of CCD imperfections for cosmology determined by weak lensing surveys:\\
From laboratory measurements to cosmological parameter bias}
\author{Yuki Okura\altaffilmark{1}\altaffilmark{2}} 
\email{yuki.okura@riken.jp}
\author{Andrea Petri\altaffilmark{3}} 
\author{Morgan May\altaffilmark{4}} 
\author{Andr\'es A. Plazas\altaffilmark{4}\altaffilmark{5}} 
\author{Toru Tamagawa\altaffilmark{1}} 
\altaffiltext{1}{RIKEN Nishina Center}
\altaffiltext{2}{RIKEN-BNL Research Center}
\altaffiltext{3}{Columbia University}
\altaffiltext{4}{Brookhaven National Laboratory}
\altaffiltext{5}{Jet Propulsion Laboratory, California Institute of Technology}

\begin{abstract}
Weak gravitational  lensing causes subtle changes in the apparent shapes of galaxies due to the bending of light by the gravity of foreground masses. By measuring the shapes of large numbers of galaxies (millions in recent surveys, up to tens of billions in future surveys) we can infer  the parameters that determine cosmology. Imperfections in the detectors used to record images of the sky can introduce changes in the apparent shape of galaxies, which in turn can bias the inferred cosmological parameters. In this paper we consider the effect of two widely discussed sensor imperfections:  tree-rings,  due to impurity gradients which cause transverse electric fields in the Charge-Coupled Devices (CCD), and pixel-size variation, due to periodic CCD fabrication errors. These imperfections can be observed when the detectors are subject to uniform illumination (flat field images). We develop methods to determine the spurious shear and convergence (due to the imperfections) from the flat-field images.  
We calculate how the spurious shear when added to the lensing shear will bias the determination of cosmological parameters. We apply our methods to candidate sensors of the Large Synoptic Survey Telescope (LSST) as a timely and important example, analyzing flat field images recorded with LSST prototype CCDs in the laboratory. We find that tree-rings and periodic pixel-size variation present in the LSST CCDs will introduce negligible bias to cosmological parameters determined from the lensing power spectrum, specifically $w$, $\Omega_m$ and $\sigma_8$.
\end{abstract}

\section{Introduction}
Weak lensing of the large scale structure (cosmic shear) is one of the most powerful methods for constraining the cosmological model and is sensitive to both the growth of structure and to the expansion history of the universe (\cite{Kaiser2000}, \cite{Wittman2000}, \cite{Miller2007}, \cite{Schneider2006}, \cite{Munshi2008}, \cite{Kilbinger2014},  \cite{Jarvis2015}). However, so far weak lensing data has been available for only small regions of the sky ($<$200 deg$^2$)

Surveys that will produce large weak lensing data sets have begun (Dark Energy Survey \footnote{http://www.darkenergysurvey.org/}(DES), Hyper Suprime-Cam \footnote{http://www.naoj.org/Projects/HSC/HSCProject.html}(HSC), Panoramic Survey Telescope And Rapid Response System(PanStarrs) \footnote{pan-starrs.ifa.hawaii.edu/}, The Kilo-Degree Survey \footnote{http://kids.strw.leidenuniv.nl/}(KiDS)) or are being built (LSST\footnote{http://www.lsst.org}, Euclid\footnote{http://sci.esa.int/euclid}, The Wide-Field Infrared Survey Telescope (WFIRST)\footnote{http://wfirst.gsfc.nasa.gov/}). The statistical power of forthcoming weak lensing data sets, which will cover thousands or tens of thousands of square degrees, makes it necessary to consider the effect of systematic errors that could previously be ignored.  The cosmic shear analysis relies on the accurate measurement of shapes of as many as tens of billions of background galaxies, to statistically estimate the small distortion (about $1\%$) caused by weak gravitational lensing. To fully benefit from the increased statistical power of the new surveys, systematic errors must not significantly degrade the cosmological parameter errors.

Some imperfections in CCDs become apparent when they record a uniform illumination (flat fields). Structure in the images reveals several patterns that are not due exclusively to quantum efficiency variations. Concentric arcs, called tree rings (\cite{plazas2014}, \cite{Stubbs2014}, \cite{Lupton2014}, \cite{Jarvis2014}, \cite{Holland2014}, \cite{Rasmussen2015}) are due to impurity gradients in the CCDs which cause electric fields transverse to the surface of the CCD. Patterns aligned with the rows and columns of the pixels are due to periodic pixel-size variation caused by errors in the masks used in the step-and-repeat process manufacturing (\cite{Smith2008}). Both these effects can change the shape of galaxy images. 
For weak lensing analyses, we quantify the shape changes as spurious shear or spurious convergence.  

For the tree rings, we show how to calculate the spurious shear and spurious convergence from the flat field images. For the pixel-size variation, we calculate the spurious convergence which is directly related to the size of the pixel.

The primary statistics for inferring cosmology from weak lensing surveys is the 2-point correlation function of shear or convergence, and its Fourier transform, the power spectrum.  The 2-point correlation function of spurious shear must be much smaller than that due to cosmological shear if cosmological parameters are to be accurately inferred. 

 The biases in cosmological parameters are determined by calculating how much the inferred parameters change when the spurious convergence is added to the lensing convergence (\cite{petri2014}). 

This paper is organized as follows: In section 2 we give an overview of how we compute the bias in cosmological parameters from the spurious convergence. In section 3, we describe the measurement of tree rings, the spurious 2-point correlation function it causes,  and the parameter bias it induces. In section 4, we describe the measurement of pixel-size variation, the 2-point correlation function of the spurious convergence and the parameter bias it induces. We summarize our findings and discuss further work in section 5.

\section{Spurious shear and spurious convergence}

Weak gravitational lensing causes changes in the apparent position of point sources. For a source at $\pmb{\theta}_S$  observed at displaced position $\pmb{\theta}$ due to lensing, the shift is described by a 2$\times$2 transformation matrix (\cite{Dodelson2003})  

\begin{equation}
A_{ij}\equiv \frac{ \partial \theta_S^i}{\partial \theta^j}
\equiv\left(
\begin{array}{cc}
1-\kappa-\gamma_1&-\gamma_2\\
-\gamma_2&1-\kappa+\gamma_1
\end{array}
\right)
\end{equation}
where $\kappa$, the convergence, describes the isotropic distortion (magnification) of an extended source, and $\gamma_1$, $\gamma_2$, the shears, describe anisotropic distortions of the image.  Detector imperfections can cause distortions of the images of galaxies resulting in spurious shear and convergence, biasing the cosmological parameter determination.

Previous work has described biases due to the atmosphere, telescope and detector in terms of spurious shear (\cite{Chang2013}). 
However the convergence and the shear are alternative descriptions of lensing, the convergence is related to shear (up to a uniform mass sheet degeneracy) through a non-local Kaiser-Squires inversion (\cite{KaiserSquires1993}). The power spectrum of convergence is equal to the power spectrum of E-mode shear. We point out here that spurious convergence is in some cases a more convenient description of CCD imperfections than spurious shear. 
The spurious convergence due to pixel-size variation is easily calculated from the change in the pixel area from its nominal value. 
Since for lensing, convergence is proportional to the mass surface density, it is clear that if we consider an area composed of many pixels, the convergence is just the average of the convergence of the constituent pixels.

\section{Bias to cosmological parameters}
The bias calculation is performed using a Fisher formalism (see for example \cite{Dodelson2003}. Let $\hat{\kappa}^0$ be the convergence from weak lensing in a field of view of size $\theta_{\rm FOV}$, and let $\hat{P}_i^0$ be its power spectrum defined as 
\begin{equation}
\langle\tilde{\kappa}(\pmb{\ell}_i)\tilde{\kappa}(\pmb{\ell}_j)\rangle = (2\pi)^2\delta_D(\pmb{\ell}_i+\pmb{\ell}_j)P_i
\end{equation}
We used the superscript 0 to denote the convergence measured from a pure lensing signal, not including systematic effects. In this equation, and in the remainder of this section, the hat ($\hat{q}$) on a quantity $q$ means that the quantity is an estimator and we use the tilde $\tilde{q}$ to denote the Fourier transform of $q$. The subscripts $i,j$ are shorthands for the multipole moments $\pmb{\ell}_i,\pmb{\ell}_j$. We also indicate the Dirac delta function and the Kronecker delta symbol as $\delta_D,\delta_{ij}$ respectively.
In the limit in which $\hat{\kappa}^0$ is a Gaussian field, the power spectrum estimator $\hat{P}^0_i$ has an expectation value $\langle\hat{P}^0_i\rangle=P_i$ and a variance $\langle(\hat{P}^0_i-P^0_i)^2\rangle=(P^{0}_i)^2/N_i$, where the mode number $N_i$ is computed as see \citep{Dodelson2003} 

\begin{equation}
N_i = \frac{\ell_i\delta\ell_{\rm bin}\theta^2_{\rm FOV}}{4\pi} = f_{\rm sky}\delta\ell_{\rm bin}\ell_i
\end{equation}
where $\delta\ell_{\rm bin}$ is the width of the multipole bins used in the analysis.  Because the convergence field is statistically isotropic, the power spectrum measured at a multipole $\pmb{\ell}$ depends on the magnitude $\ell$ of the two dimensional multipole moment $\pmb{\ell}$. Now let $X_{\alpha i}$ the derivative of $P^0_i$ with respect to cosmological parameter $\alpha$ and $C_{ij}$ the $\hat{P}^0_i$ estimator covariance matrix, which we approximate as diagonal 
\begin{equation}
C_{ij} = \langle(\hat{P}^0_i-P^0_i)(\hat{P}^0_j-P^0_j)\rangle = \delta_{ij}\frac{(P^0_i)^2}{N_i}
\end{equation}
Defining

\begin{equation}
e^{\rm FOV}_\alpha = \sqrt{\sum_{ij}M_{\alpha i}M_{\alpha j}C_{ij}}
\end{equation}
we can estimate the bias $b_\alpha$ in the cosmological parameters $p_\alpha$ when fitting a field of view with spurious convergence added $\hat{\kappa}^{0+sp}=\hat{\kappa}^{0}+\kappa^{sp}$ as 
\begin{equation}
\label{biasexpr}
\hat{b}_{\alpha} = M_{\alpha i} (\hat{P}^{0+sp}_i - \hat{P}^0_i).
\end{equation}
where the superscripts $0+sp$, 0 refer to the power spectrum measured from a field of view with and without spurious convergence added (\cite{petri2014}). The same Fisher formalism can be used to estimate the marginalized errors $e_\alpha$ on the parameters when using a single field of view
\begin{equation}
\mathbf{e}_{\rm FOV} = \sqrt{\mathrm{diag}(M_{\alpha i}M_{\beta j}C_{ij})}
\end{equation}
We note that when using the full survey data rather than a single field of view, parameter errors can be much smaller since $\mathbf{e}_{\rm survey} = \mathbf{e}_{\rm FOV}/\sqrt{N_{\rm FOV}}$. For LSST, $\theta_{\rm FOV}=3 {\rm deg}$  and $N_{\rm FOV} = \Omega_{\rm survey}/\theta^2_{\rm FOV}\approx 2000$. 

Equation (\ref{biasexpr}) can be further expanded as  

\begin{equation}
\label{biasestimator} 
\hat{b}_{\alpha} = M_{\alpha i}\left[\frac{\vert{\hat\tilde{\kappa}}^0_i+\kappa^{sp}_i\vert^2}{(2\pi)^2} - \hat{P}^0_i\right] =  M_{\alpha i}\left(P^{sp}_i + \frac{\hat{\tilde{\kappa}}_i^0\tilde{\kappa}_i^{sp,*}+\hat{\tilde{\kappa}}_i^{0,*}\tilde{\kappa}_i^{sp}}{(2\pi)^2}\right)
\end{equation}
The spurious convergence $\kappa_i^{sp}$ is calculated from laboratory measurement of the CCDs. Since the imperfections are fixed on the CCD focal plane, there is only one realization of the spurious convergence and it spans the field of view of the telescope.  Because of this, the spurious convergence $\kappa^{\rm sp}$ becomes a non stochastic quantity (and hence written without the hat). We can compute the expectation value and standard deviation of the bias estimator (\ref{biasestimator}) and get 
\begin{equation}
\label{biasmean}
b_\alpha = \langle\hat{b}_\alpha\rangle = M_{\alpha i}P^{sp}_i
\end{equation}

\begin{equation}
\label{varbias}
\Delta b_\alpha = \sqrt{\langle(\hat{b}_\alpha-b_{\alpha})^2\rangle} = 2M_{\alpha i}\sqrt{\frac{P^{sp}_iP^0_i}{N_i}}
\end{equation}
Since the spurious convergence is non stochastic, the standard deviation of the bias estimator (\ref{varbias}) is proportional to the standard deviation of $\hat{\tilde{\kappa}}_i^0$, which scales as $1/\sqrt{N_i}$.
These results are correct for a single field of view, when the analysis is scaled to the full survey side the expectation value of the bias does not change, but  the standard deviation of $\hat{\tilde{\kappa}}_i^0$ gets reduced by a factor of $\sqrt{N_{\rm FOV}}$. The same reduction applies to the bias variance in (\ref{varbias}).
We note that the stochasticity in the bias is due to the cross term in (\ref{biasestimator}) and is a result of cosmic variance in the lensing convergence field. $\Delta b_\alpha$ can be quite large compared to $b_{\alpha}$ even for $N_{\rm FOV}\approx 2000$. The estimated survey bias is

\begin{equation}
\label{survey bias}
\mathbf{b}_{\rm survey }=b_\alpha \pm \Delta b_\alpha 
\end{equation}.

We will quote the larger of the two values as the estimated survey bias.

To compute $P_i$ in the fiducial cosmological model with $(\Omega_m,w,\sigma_8)=(0.26,-1,0.8)$, as well as the derivatives $X_{\alpha i}$, we make use of the public code NICAEA (\cite{Kilbinger2009}) which gives good accuracy for the lensing power spectra in the parameter range we are considering.  We assume a redshift distribution of galaxies concentrated in single redshift $z_s=2$. In the lensing convergence power spectrum $\hat{P}^0_i$ we include the effects of galaxy shape for the assumed $n_g=30$ galaxies/$\rm{arcmin}^2$.  An approximate scaling relation of the bias with the galaxy density can be obtained noting that, in the limit in which shape noise dominates, $P_0^i$ scales as $1/n_g$. Because the derivatives $X$ and the spurious power $P_i^{\rm sp}$ do not depend on $n_g$, the bias expectation value does not change with $n_g$, but its standard deviation roughly scales with $1/\sqrt{n_g}$.

\section{Tree rings}
Tree rings are due to impurity gradients in the silicon of which CCDs are made. The high-purity, high-resistivity silicon used in recent astronomical CCDs grows cylindrically from molten state. Time variations in temperature, composition, etc. produce radial impurity gradients which result in resistivity gradients. CCDs made from slices of the cylindrical boule have electric fields transverse to the main field due to the resistivity gradients leading to the displacement of photo-generated charge. Thick CCDs used for increased near IR  ($<$ 1 micron) sensitivity suffer greater charge displacement due to longer path length.

Though the tree rings are observed in flat field images, they do not correspond to quantum efficiency variations.
Instead, they induce a displacement of the collected charge that propagates into astrometric and photometric biases. 
In this section we describe how to calculate the spurious shear and spurious convergence due to tree rings from the flat-field images and apply the method to the LSST candidate sensors. We then calculate the 2PCF and bias of cosmological parameters. 
\subsection{Displacement caused by tree rings}
In practice, it is possible to directly measure the astrometric displacement $d(r)$ as function of scalar radius $r$, caused by the tree rings by using sets of several dithered exposures of star fields in different photometric bands and constructing ``star flat" images (\cite{Manfred1995}, \cite{Tucker2007}). The displacement is defined from differences between the original position $r_o$ and the displaced position by the tree-ring effect $r_d$; it is defined as
\begin{eqnarray}
\label{eq:TR_d}
d(r_d)=r_d-r_o.
\end{eqnarray}

However, this is not an easy task given that the displacement is small (of the order of subpixels).
However, it is easy to measure flux modulation $f(r)$ due to tree-ring, defined as
\begin{eqnarray}
\label{eq:TR_deff}
f(r)\equiv\frac{F(r)-F_{all}}{F_{all}}
\end{eqnarray}
where $F(r)$ is average flux at radius $r$ and $F_{all}$ is average flux of all pixels.
 \cite{plazas2014} demonstrated that, to first order, there exists a relationship between the tree-rings flux modulation $f(r)$ as measured by the flat fields (which have higher S/N compared to the one provided by the limited number of stars in the star flats) 
and the astrometric displacement $d(r)$, given by 
\begin{eqnarray}
\label{eq:TR_d_f}
d(r_d)=-\frac{1}{r_d}\int_0^{r_d} dr rf(r).
\end{eqnarray}
Note the sign convention here differs from that in \cite{plazas2014}.

\subsection{Shear and convergence caused by tree rings}
\begin{figure}[tbp]
\centering
\includegraphics[width=.5\textwidth]{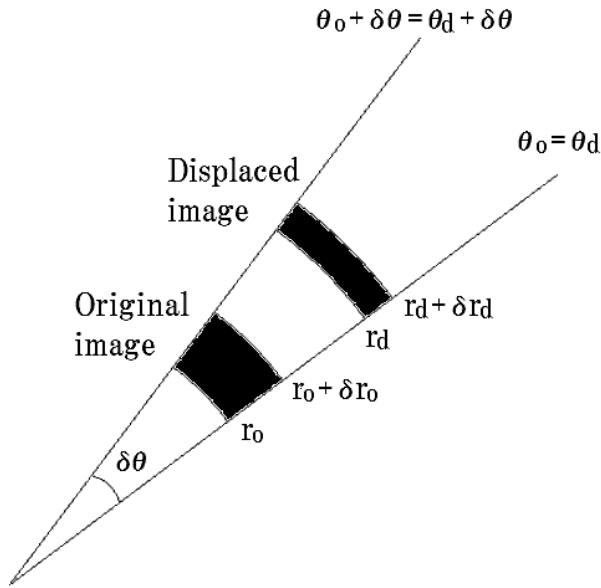}
\caption{Shape change of an area caused by  the radial displacement field  $d(r)$.}
\label{fig:CD}
\end{figure}

Because the displacements caused by tree rings have only radial components, shape changes are measured easily by considering the change in radial length and tangential length of images. Consider the effect of the displacement field on an infinitesimal image between radius $r_d$ and $r_d+\delta r_d$ and $\theta_d$ and $\theta_d+\delta \theta_d$ in polar coordinates, where the origin of the coordinate system is set to the center of the concentric displacement. Thus, the image is mapped from the original position $r_o$ and $r_o+\delta r_o$ and $\theta_o$ and $\theta_o+\delta \theta_o$ by the tree ring displacement field, which is written as a function $d(r_d)$. Figure \ref{fig:CD} shows the relation between the images. The extent of the image in the $r$ direction relate to each other  as in equation \ref{eq:TR_d} and
\begin{eqnarray}
\delta r_o\approx \delta r_d\lr{1-\frac{\partial d(r)}{\partial r} \bigg |_{r=r_d }}.
\end{eqnarray}
The shape change can be obtained by taking the ratio of the length of the images in both the radial and tangential direction. Thus, the radial size change is given by
\begin{eqnarray}
\label{eq:cdist_r}
\frac{\delta r_o }{\delta r_d}\approx1-\frac{\partial d(r)}{\partial r}  \bigg |_{r=r_d},
\end{eqnarray}
and the tangential size change is given by
\begin{eqnarray}
\label{eq:cdist_t}
\frac{r_o\delta \theta}{r_d\delta \theta}=1-\frac{d(r_d)}{r_d}.
\end{eqnarray}
For convenience, and without loss of generality, let us consider a situation where $\theta_o=0$. Then we can arrange equation \ref{eq:cdist_r} and \ref{eq:cdist_t} in matrix form, obtaining
\begin{equation}
\left(
\begin{array}{c}
\delta r_o\\
r_o\delta\theta
\end{array}
\right)=\left(
\begin{array}{cc}
1-\frac{\partial d(r)}{\partial r}|_{r=r_d}&0\\
0&1-\frac{d(r_d)}{r_d}
\end{array}
\right)\left(
\begin{array}{c}
\delta r_d\\
r_d\delta\theta
\end{array}
\right) \\ 
\equiv\left(
\begin{array}{cc}
1-\kappa^{TR}-\gamma_{rad}^{TR}&0\\
0&1-\kappa^{TR}+\gamma_{rad}^{TR}
\end{array}
\right)\left(
\begin{array}{c}
\delta r_d\\
r_d\delta\theta
\end{array}
\right)
\end{equation}
where $\kappa^{TR}$ and $\gamma^{TR}$ are the convergence and shear due to the tree-rings displacement, defined as
\begin{eqnarray}
\label{eq:TR_conv_d}
\kappa^{TR}(r)&\equiv&\frac12\lr{\frac{\partial d(r)}{\partial r}+\frac{d(r)}{r}}\\
\label{eq:TR_shear_d}
\gamma^{TR}_{rad}(r)&\equiv&\frac12\lr{\frac{\partial d(r)}{\partial r}-\frac{d(r)}{r}},
\end{eqnarray}
where $\gamma_{rad}$ is positive for radial shear and negative for tangential shear.
Thus, the two components of the shear, $\gamma_1$ and $\gamma_2$, at $(r,\theta)$ can be written as
\begin{eqnarray}
\label{eq:TR_shear_12}
\gamma^{TR}_{1}(r,\theta)+i\gamma^{TR}_{2}(r,\theta)&=&\gamma^{TR}_{rad}(r)e^{2i\theta}
\end{eqnarray}
By using equation \ref{eq:TR_d_f}, \ref{eq:TR_conv_d} and \ref{eq:TR_shear_d} can be written in terms of the flux modulation
\begin{eqnarray}
\label{eq:TR_conv_f}
\kappa^{TR}(r)&=&-\frac12f(r)\\
\label{eq:TR_shear_f}
\gamma^{TR}_{rad}(r)&=&-\frac12\lr{f(r)-2\frac{d(r)}{r}}\approx-\frac12f(r),
\end{eqnarray}
where the last approximation is valid when the displacement is much smaller than the radius. The typical spatial scales of the displacements are of the order of subpixels, and the radii of the order of 100$\sim$1000 pixels, so the approximation is valid. 
This means that the convergence and shear due to the tree rings are approximately half the value of the flux modulation.

For shear $\gamma_1 + i \gamma_2 = \gamma e^{i2(\theta_\gamma)}$  at $(r,\theta)$  in polar coordinates relative to a fixed origin, the radial and cross shear are given by  $\gamma_{rad} + i \gamma_{\times} = \gamma e^{i2(\theta_\gamma-\theta)}$. The tangential shear is obtained as $\gamma_{t}=-\gamma_{rad}$.

\subsection{Measurement of tree rings on LSST CCDs}
We measured the tree-ring patterns of two different LSST candidate CCDs.  Here we discuss only one CCD type, since results for the other were similar. We used flat-field images from a uniform light source of $750$ nm wavelength. The CCDs have approximately 4000$\times$4000 pixels, each one with a physical size of 10 microns $\times$ 10 microns.  The CCDs are 100 microns thick for near IR sensitivity. 
We have 25 flat field images, and each image has approximately 62500 mean electron counts per pixel.
First we corrected the flat field images to eliminate other effects by following steps:
\begin{itemize}
\item Masked regions near the edge of the detectors and shadows due to dust particles in the optics.
\item The light source in this test was not completely uniform, so fit the slow variation with a 7th order polynomial for each readout channels, then normalized by the polynomial.
\item Stacked the normalized 25 flat images.
\item Measured the laser annealing pattern and pixel variation pattern, then simply divided the flat fields by the patterns.
\item Removed other large scale fluctuation in Fourier space.
\end{itemize}
Though pixel-size variation also will result in its own pattern of spurious distortion, for the tree-ring analysis we removed 
both pixel-size variation and  laser annealing effects by scaling the image. We discuss pixel-size variation in detail in the next section.

\begin{figure}[tbp]
\centering
\includegraphics[width=\textwidth]{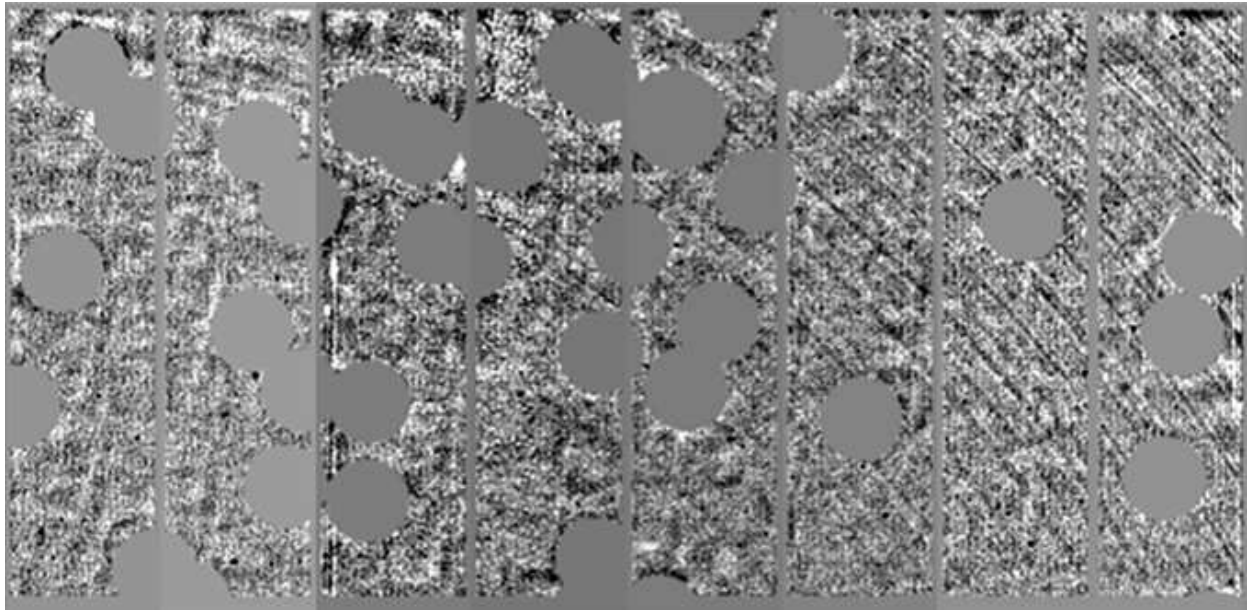}
\caption{Corrected flat-field image with smoothing of half of upper region of one LSST prototype CCD. The masked regions are shown in uniform gray.
}
\label{fig:SSTR_2dim}
\end{figure}

We determine the center of the tree ring profiles in each CCD by selecting points on several rings and then fitting for their common center. The center of the rings  lies  near of the corner of the CCD, but slightly offset from it (see Figure \ref{fig:SSTR_2dim}). 

\begin{figure}[tbp]
\centering
\includegraphics[width=0.95\textwidth]{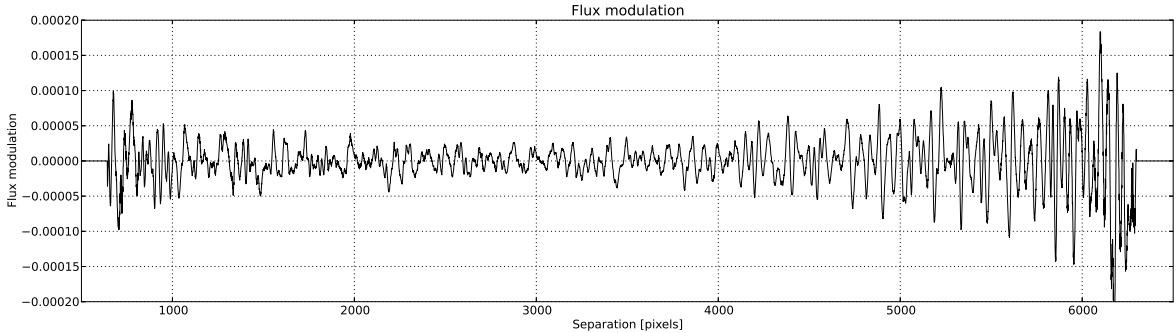}
\caption{1-dimensional profile of the flux modulation (adimensional) caused by tree rings in LSST CCDs. The spurious convergence and shear are each 1/2  the flux modulation (see Equations 15 and 16).
}
\label{fig:SSTR_1dim}
\end{figure}

Figure \ref{fig:SSTR_1dim} shows the measured tree-ring profiles. 
Before measuring the profiles, the images were smoothed by a Gaussian kernel with a 5-pixel standard deviation. Since the width scale of the tree rings peaks is about $100$ pixels, the smoothing should not reduce their amplitude by much. The typical amplitude of the flux modulation is approximately $0.01\%$, about 50 times smaller than that on the DECam CCDs reported by \cite{plazas2014}. The tangential spurious shear is calculated from the flux modulations by using Equation \ref{eq:TR_shear_f} (the spurious shear is about half value of flux modulation). The typical amplitude of the spurious shear is of about $0.005$\%.
The displacement is calculated by using Equation  \ref{eq:TR_d_f}. 
The typical amplitude of the astrometric displacement is about $0.001$ pixels $\sim 0.0002$ arcseconds. This means the displacement caused by tree-ring effect does not cause significant missregistrations.

\subsection{2-point correlation function of tree-ring spurious shear}
\begin{figure}[tbp] 
\centering
\includegraphics[width=\textwidth]{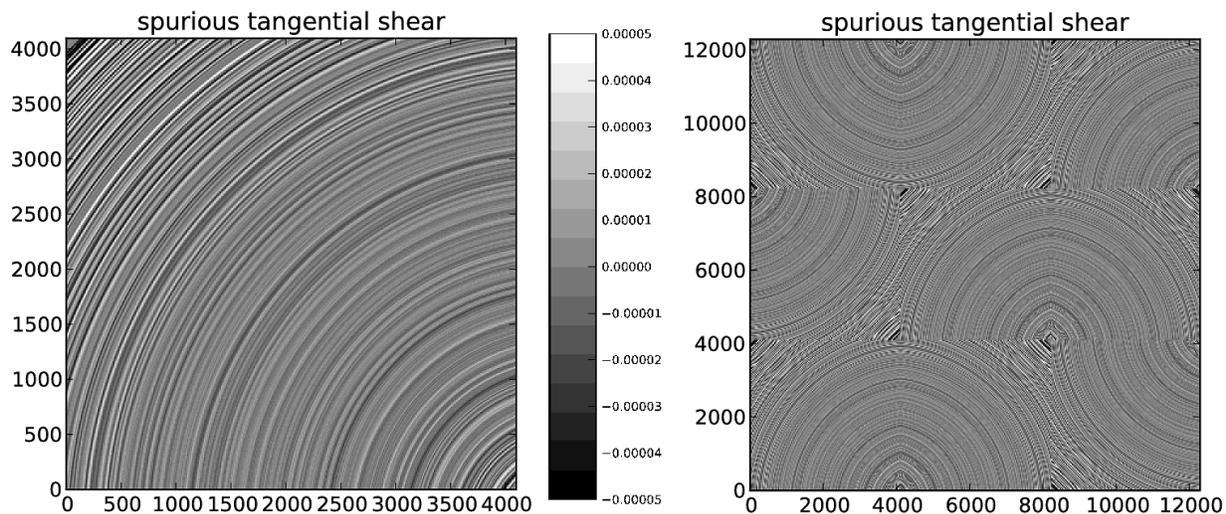}
\caption{Tangential spurious shear produced by the tree rings on an LSST CCD (left panel) and on a 9-CCD raft (right panel), calculated from the flux modulation measured in the flat fields. The raft is composed of CCDS having the same tree-ring profiles but with random orientations
differing by 90 degree rotations.}
\label{fig:SSTR_2dim_color}
\end{figure}

Here we calculate the 2-point correlation function (2PCF) of the spurious shear induced by tree rings on the LSST CCDs.  First, we calculate the spurious shear pattern on each individual CCD from the observed flux modulation (Figure \ref{fig:SSTR_2dim_color} left panel). The spurious shear 2PCF is calculated from $\gamma_t$ and $ \gamma_\times$ (see for example \cite{Kilbinger2014})

\begin{equation}
\label{2pcf}
\xi_+(\theta)=\lrt{\gamma_t \gamma_t}(\theta) +\lrt{\gamma_\times \gamma_\times}(\theta).
\end{equation}

A typical galaxy used for LSST weak lensing analysis subtends 1 arcsec, comparable to the 0.7 arcsec PSF. Each pixel subtends 0.2 arcsec. So a typical galaxy image covers $\approx5\times5$ pixels.
The scale on which  tree-ring shear varies is approximately 100 pixels, so the spurious shear does not change significantly over a galaxy image. Thus the spurious shear for a galaxy is approximately equal to the spurious shear at any pixel within the galaxy image.

The LSST focal plane is composed of 21 rafts, each raft composed of 9 CCDs. We assume that all CCDs in the raft have the same tree ring profile but with random orientations differing by 90 degree rotations. The right panel of Figure  \ref{fig:SSTR_2dim_color} shows the spurious shear pattern for a single raft. We assemble a virtual LSST focal plane out of 25 rafts arranged in square applying random 90 degree rotations to each raft.

\begin{figure}[tbp] 
\centering
\includegraphics[width=\textwidth]{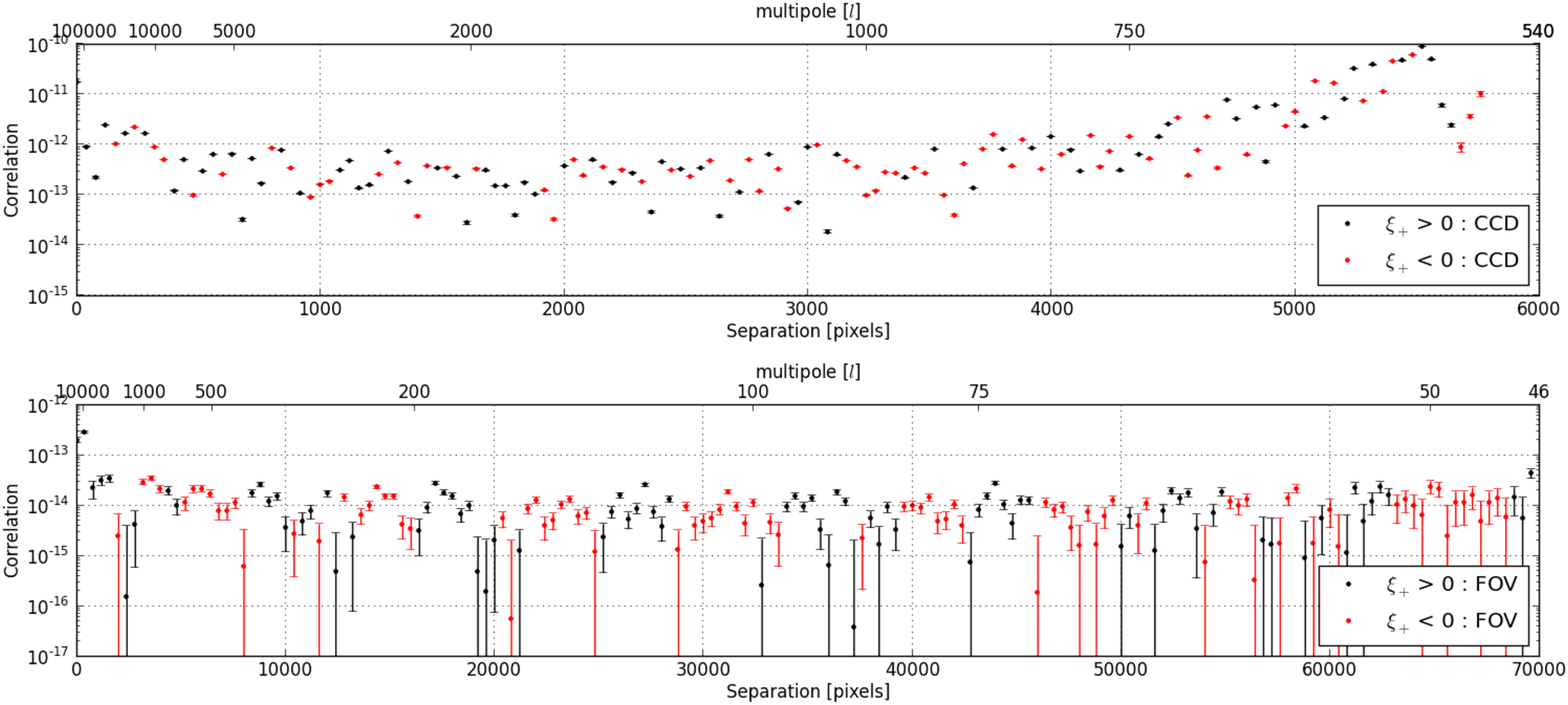}
\caption{The 2-point correlation function $\xi_+$ of spurious shear caused by  tree rings  on LSST CCDs. Absolute values are shown; the values oscillate around zero. Upper panel is for a single CCD. Lower panel is for the full LSST focal plane. 
}
\label{fig:SSTR_2cor_CCD_log}
\end{figure}

Figure  \ref{fig:SSTR_2cor_CCD_log} upper panel shows the 2PCF of spurious shear due to tree rings for a single LSST CCD. The typical value is about $10^{-12}$.  At very short spatial scales we can see a slightly larger peak. This value is much smaller than the square of the typical amplitude of the spurious shear, $(5\times10^{-5})^2 \approx 10^{-9}$. 

The smallness of the 2PCF results from a cancellation between alternate signs of spurious shear in its calculation. The 2PCF averages over pairs of galaxies at a fixed separation. The sign of  the spurious shear due to tree rings alternates on the tree ring scale as is seen in Figure  \ref{fig:SSTR_2dim_color}. With 30 galaxies per square arcmin, each CCD will record $\approx$5$\times 10^3$ galaxies per pointing, and $\approx 10^7$ galaxies for the  $\approx$2$\times 10^3$ pointing over the course of the survey. 
Dithering will further reduce the effect of the imperfections we consider.
Thus there are a huge number of pairs of galaxies at similar separations but with differing sign of spurious shear.  There is less of a cancellation on scales comparable to the CCD size, due to the smaller number of pairs at that separation. At scales less than the 100 pixel tree ring scale, both points may lie on the same tree ring.


Fig. \ref{fig:SSTR_2cor_CCD_log} lower panel shows the 2PCF of the spurious shear on an area equivalent to the full field of view of the LSST camera with a $100$ pixel sampling scale. The typical scale for the amplitude decreases to $10^{-14}$ due to further averaging down for the reasons described above.

In comparison, the lensing 2PCF $\xi_+$ predicted by NICAEA (\cite{Kilbinger2009}) for the fiducial cosmology with a LSST-like galaxy distribution varies monotonically from  $2\times 10^{-4 }$ at 1 arcsec separation (5 pixels ) to $5\times 10^{-7}$ at $10^4$ arcsec separation (20,000 pixels).

The fact that the spurious shear two point function is much smaller than the weak lensing signal is encouraging, but is not sufficient by itself to say that the bias on the cosmological parameters is negligible. To quantify the former bias, we need to isolate the features of the two point function that are most sensitive to cosmology, and study how the spurious power in this feature space affects our parameter estimates. We quantify the bias on the cosmological parameters in the following section.

\subsection{Bias in the cosmological parameters caused by tree rings}
\begin{figure*}[htbp]
\centering
\resizebox{0.6\hsize}{!}{\includegraphics{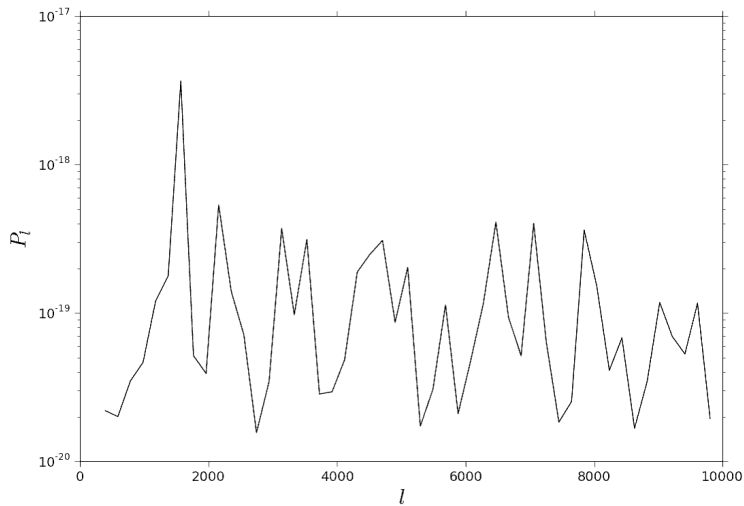}}
\caption{
\label{fig:Pl_TR}
Power spectrum of spurious convergence caused by  tree rings.
}
\end{figure*}

The bias in cosmological parameters caused by tree rings can be calculated from the spurious convergence as described in Section 3. 
Figure \ref{fig:Pl_TR} shows the spurious convergence power spectrum  $P_l$ used to calculate the bias caused by tree rings. The spurious power fluctuates randomly around $10^{-19}$ and does not show a systematic trend with $l$. The spurious power spectrum was calculated on a grid with each element of the grid 30$\times$30 pixels and the convergence averaged over the grid element.

Table \ref{tab:Bias_TR} shows the amplitude of the bias due to tree rings.
The biases shown in Table \ref{tab:Bias_TR} are very much smaller than the expected marginalized errors $\mathbf{e}$ for LSST which are are 0.00232 for $\Omega_m$,  0.02434 for $w$ and  0.00427  for $\sigma_8$, calculated as described in Section 3 assuming 30 galaxies per square arcminute at z=2. The marginalized errors will be somewhat smaller if tomography is used.

\begin{table}[htbp]
\centering
\begin{tabular}{|c|c|c|c|}\hline
&$\Omega_m$&$w$&$\sigma_8$\\\hline
bias expectation value	& 5.05E-10	& 2.79E-09 &	-3.52E-10\\\hline
bias standard deviation	& 6.92E-08 &	1.34E-07	& 1.29E-07\\\hline 
\textbf{Survey bias }    & 6.97E-08  &	1.36E-07 &	1.29E-07\\\hline 
\end{tabular}
\caption{\label{tab:Bias_TR}
Bias of the cosmological parameters ($\Omega_m, w, \sigma_8$) caused by tree rings. 
}
\end{table}

\section{Periodic pixel-size variation}

Patterns aligned with the rows and columns of the pixels are due to periodic pixel-size variation caused by errors in the masks used in CCD fabrication. We do not consider here aperiodic pixel-size variation which is degenerate with quantum efficiency variations and hence difficult to measure (\cite{Stubbs2014}).  
Aperiodic pixel-size variation is less likely to result in spurious shear-shear correlations.

In this section we describe how to calculate the spurious shear and spurious convergence due to pixel-size variation from flat field images and apply the method to the LSST candidate sensors. We then calculate the 2PCF and bias of cosmological parameters.
\subsection{Measuring periodic pixel-size variation }

In our study of pixel-size variation, we use flat field images after correction for other effects, and assume that remaining differences in counts are due only differences in pixel size. We look for periodic variation in the width of the columns and the height of rows of pixels in the CCD. We call the coordinate along direction of the rows x (horizontal) and along the direction of the columns y (vertical). The column width is assumed to be a function of x only and the row height a function of y only. The pixel boundaries are then parallel lines in the x and y direction. We find in the following analysis that the spacing between the pixel boundaries has a periodic variation consistent with the pattern observed visually in the flat fields.

Our measured quantities are the counts in each pixel in the flat fields. The width of a column is assumed to be proportional to the sum of the counts in the pixels in the column, which we call $C_c(x)$. We call the mean value of $C_c(x)$ averaged over all columns $\lrt{C_c}$. We call $\delta_h(x)$ the fractional deviation of the column width from the mean, where the subscript h is used to denote the horizontal dimension.

\begin{equation}
\label{VhC}
\delta_h(x)=\frac{C_c(x)}{\lrt{C_c}}-1
\end{equation}

Similarly, the height of a row is assumed to be proportional to the sum of the counts in the pixels in the row, which we call $C_r(y)$. We call the mean value of $C_r(y)$ averaged over all rows $\lrt{C_r}$. We call $\delta_v(y)$ the fractional deviation of the row height from the mean, where the subscript v is used to denote the vertical dimension.

\begin{equation}
\label{VvC}
\delta_v(y)=\frac{C_r(y)}{\lrt{C_r}}-1
\end{equation}

\subsection{Pixel-size variation in the LSST CCDs}
We used same flat field images to measure pixel-size variation as we used to measure tree rings. Since the effect is at the pixel scale we did not smooth the image.

Each LSST CCD is divided into 16 segments, approximately 2000x500 pixels each, each with its own amplifier for the purpose of rapid readout. The 8 channels on the lower half of the CCD are called group A, the eight channels on the upper half are called group B. The direction of the x-axis is inverted between group A and B.

\begin{figure*}[htbp]
\centering
 \resizebox{\hsize}{!}{\includegraphics{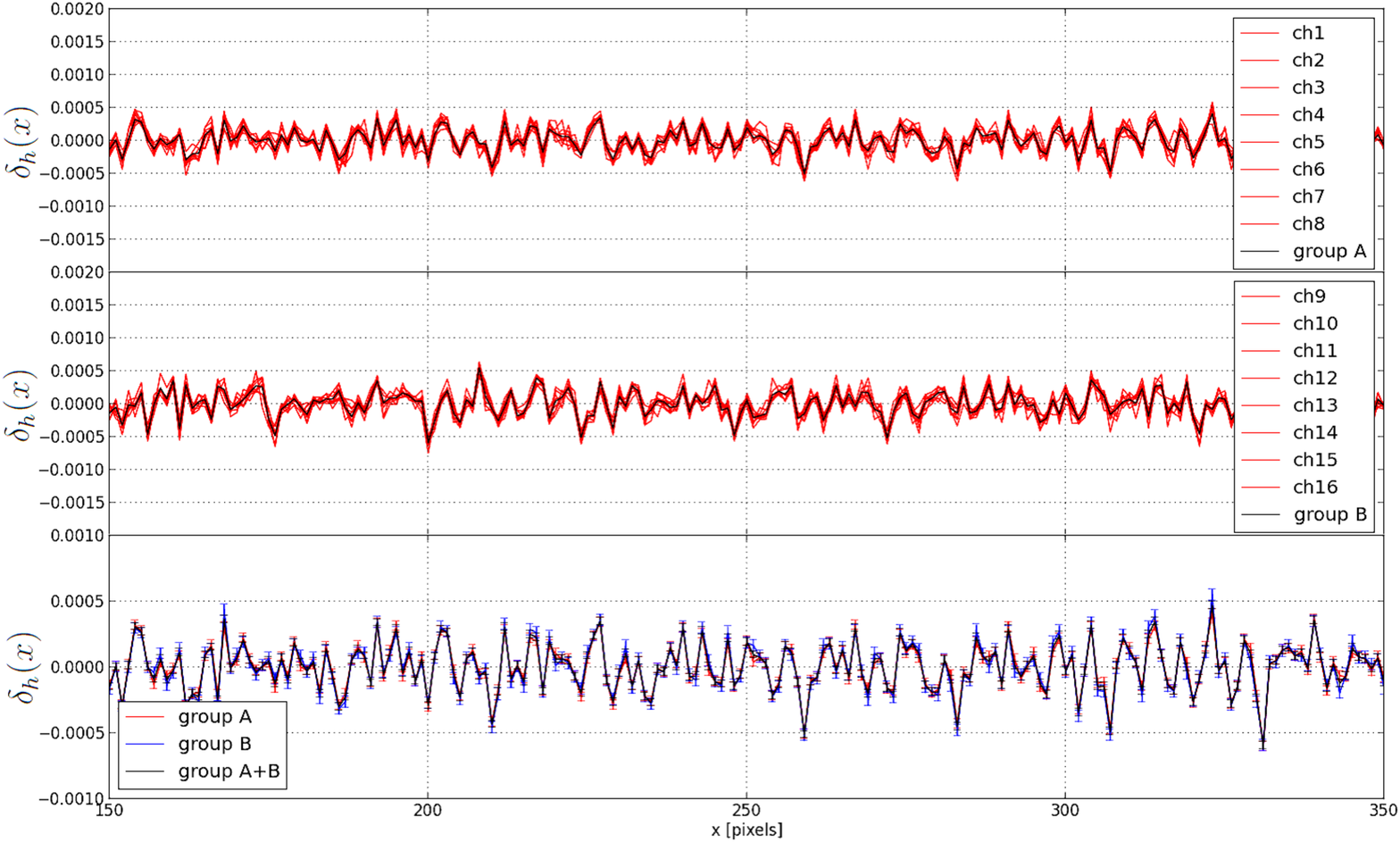}}
\caption{
\label{fig:SSPV_1dimpixelsize_chA_x}
Upper and middle panels: horizontal pixel-size variation for channels 1 to 8 (A is the average of channels 1 to 8) and channels 9 to 16 (B is the average of channels 9 to 16). $\delta_h(x)$ the fractional deviation of the column width from the mean.
Lower panel: Horizontal pixel-size variation for group A (red) and B (blue) with the horizontal  axis for  B  transformed to $x'=531-x$.  
Black line represents average horizontal pixel-size variation for all channels. The figure shows 200 out of the full 500  pixel horizontal range for each channel.
}
\end{figure*}

The red lines in the upper and middle panels of Figure \ref{fig:SSPV_1dimpixelsize_chA_x}  show the horizontal pixel-size variation in channels that belong to group A (channels 1 to 8) and group B (channels 9 to 16). Black lines in the upper (middle) panel of the figure are the average of channels in group A (B). 

In the lower panel of the figure, we plot A (red) and B (blue) superimposed and showing error bars, but with the horizontal  axis for  B  transformed to $x'=531-x$. With this transformation, group A and B have very similar profiles, and we average them to get the black curve in the lower panel; we use this average for calculating spurious shear in the following sections.

\begin{figure*}[htbp]
\centering
 \resizebox{\hsize}{!}{\includegraphics{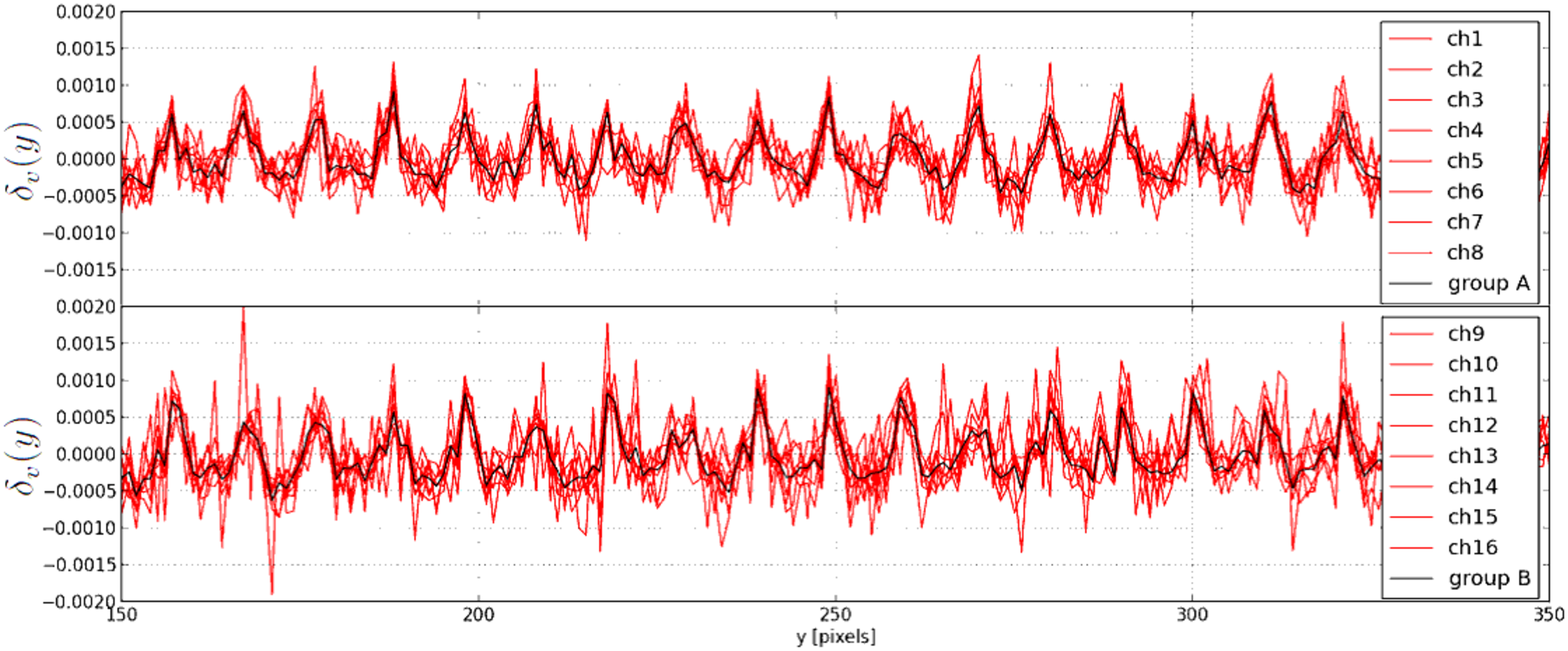}}
\caption{
\label{fig:SSPV_1dimpixelsize_chA_y}
Vertical pixel-size variation for channel 1 to 8 (group A is the mean of channels 1 to 8) and for channel 9 to 16 (group B is the mean of channels 9 to 16). $\delta_v(y)$ is the fractional deviation of the row height from the mean. The figure shows 200 out of the full 2000 vertical pixel range of a channel.
}
\end{figure*}

The red lines in Figure \ref{fig:SSPV_1dimpixelsize_chA_y}  show the vertical pixel-size variation in channels that belong to group A and B.
We can see the profiles of all channels  are similar; black lines  in the upper (lower) panel are the average of channels in group A (B). 
We can see that the vertical pixel-size variation is slightly larger than the horizontal pixel-size variation. The figure shows 200 out of the full 2000 vertical pixel range of a channel. We look for periodicity by comparing the pixel pattern with itself shifted by N pixels. In this way we find 256-pixel and 41-pixel cycles.

\begin{figure*}[htbp]
\centering
 \resizebox{\hsize}{!}{\includegraphics{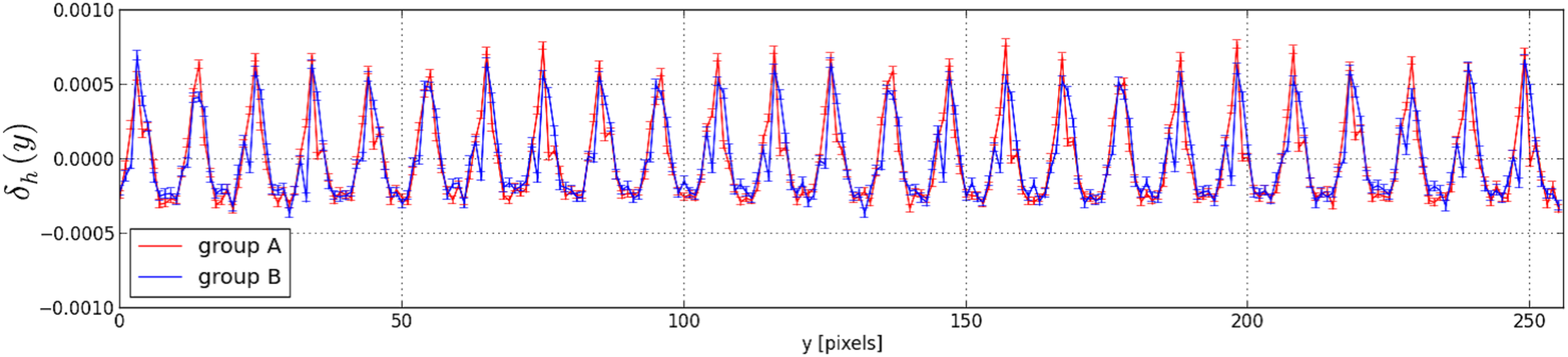}}
\caption{
\label{fig:SSPV_1dimpixelsize_chAB_256_y}
Vertical pixel-size variation for group A and B, 256-pixel cycle average.
}
\end{figure*}

\begin{figure*}[htbp]
\centering
 \resizebox{\hsize}{!}{\includegraphics{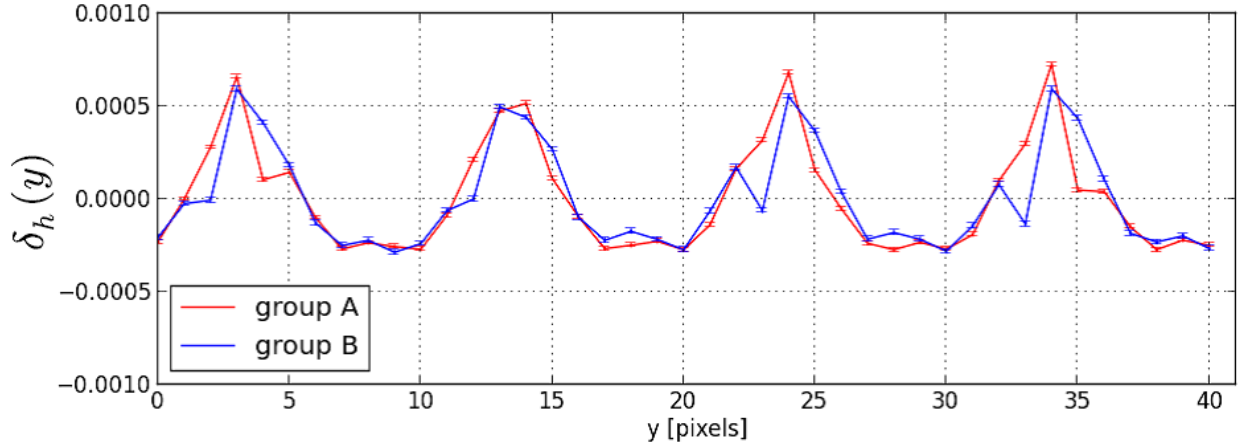}}
\caption{
\label{fig:SSPV_1dimpixelsize_chAB_41_y}
Vertical pixel-size variation for group A and B, 41-pixel cycle average.
}
\end{figure*}

The pattern of periodic peaks in Figure \ref{fig:SSPV_1dimpixelsize_chA_y}  repeats with a 256-pixels cycle.
Figure \ref{fig:SSPV_1dimpixelsize_chAB_256_y} shows the average of the repeating 256-pixels cycle.
Within the 256-pixel cycle there is periodic pattern that repeats with a 41-pixel cycle. Figure \ref{fig:SSPV_1dimpixelsize_chAB_41_y} shows the average of the repeating 41-pixel cycle. 
The 41-pixel cycle contains 4 peaks, one of the peaks spans 11 pixels, the other 3 spans 10 pixels, so the minimum pattern unit is 41 cycles.
We use the average of the 41-pixel cycle for calculating spurious convergence in following sections.

\subsection{2-point correlation functions due to pixel-size variation}
We calculated the 2 point correlation making the following  assumptions:

1) Galaxy size: The scale of pixel-size variation is about 10 pixels,  the same order as the images of galaxies (in contrast, tree-ring scale is much larger than the size of galaxy images). This means the effect of spurious shear caused by pixel-size variation should be averaged over the galaxy profile. As discussed previously, a typical galaxy image covers 5$\times$5 pixels on the LSST CCDs. Therefore, in this calculation we use a grid with each element of the grid 5$\times$5 pixels and the spurious convergence averaged over the grid element. 

2) Random position for objects: If the positions used to calculate the 2PCF are aligned with the grid, the 2PCF may be inappropriately enhanced.

3) PSF effect ignored: The spurious shear caused by pixel-size variation affects the galaxy image after passing through the atmosphere. In most cases, the PSF correction augments the spurious shear caused by pixel-size variation, e.g. in the situation of a circular PSF. 
As typical situation, let us consider an elliptical Gaussian galaxy and circular Gaussian PSF with same sizes, where the PSF smears the galaxy and reduces its ellipticity by a factor of 2.
Pixel size variation in the CCD will then add additional ellipticity to the smeared galaxy.
The gain is different for each PSF and galaxy size.
When the PSF correction is made, we recover the original ellipticiy of the galaxy plus twice the ellipticity induced by pixel size variation.

4) The spurious shear can affect the PSF correction itself. However, since the typical distance between stars used for the PSF correction is much greater than spatial scale of variation of the spurious shear, the effect is random, and is averaged down in the interpolation. 
Note that this is also true in the case of tree rings.

\begin{figure*}[htbp]
\centering
\includegraphics[width=\textwidth]{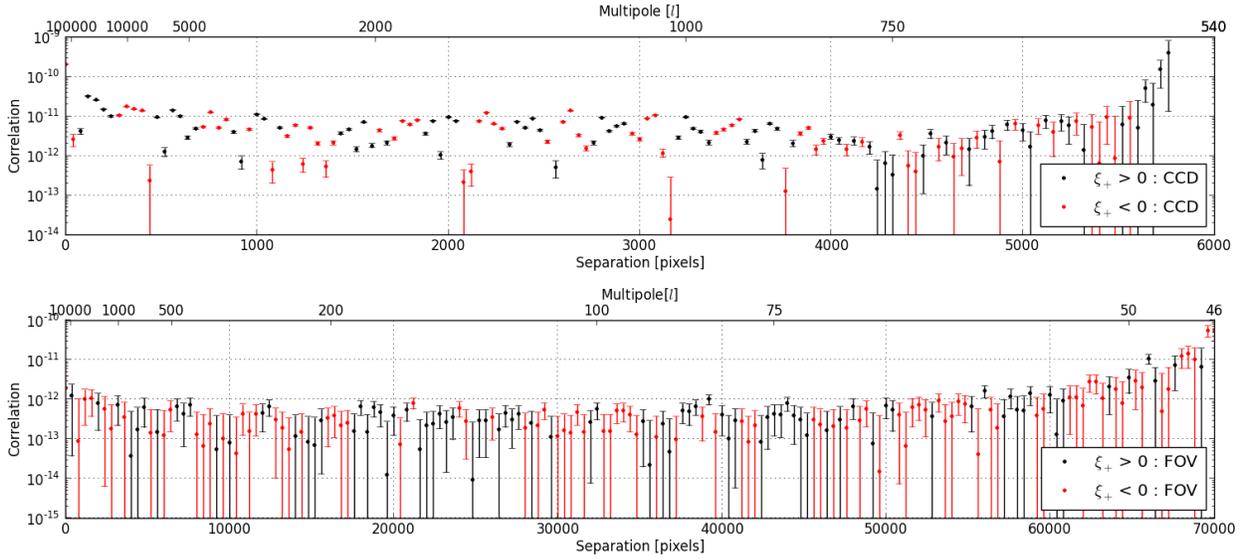}
\caption{
The 2-point correlation function $\xi_+$ of spurious convergence caused by pixel-size variation for LSST CCDs. Absolute values are shown; the values oscillate around zero. Upper panel is for a single CCD. Lower panel is for the full LSST focal plane.
}
\label{fig:SSPV_2cor_CCD_log}
\end{figure*}

Figure \ref{fig:SSPV_2cor_CCD_log} shows the absolute value of 2-point correlation of spurious convergence due to pixel-size variation calculated with $10^6$ randomly positioned points, taking into account all possible pairs. In the upper panel (single CCD) the typical scale of the 2PCF is $10^{-10}$. In the lower panel (full LSST focal plane) the typical scale of the 2PCF is $10^{-11}$.

\clearpage
\subsection{Bias in the cosmological parameters caused by pixel-size variation}
\begin{figure*}[htbp]
\centering
\resizebox{.6\hsize}{!}{\includegraphics{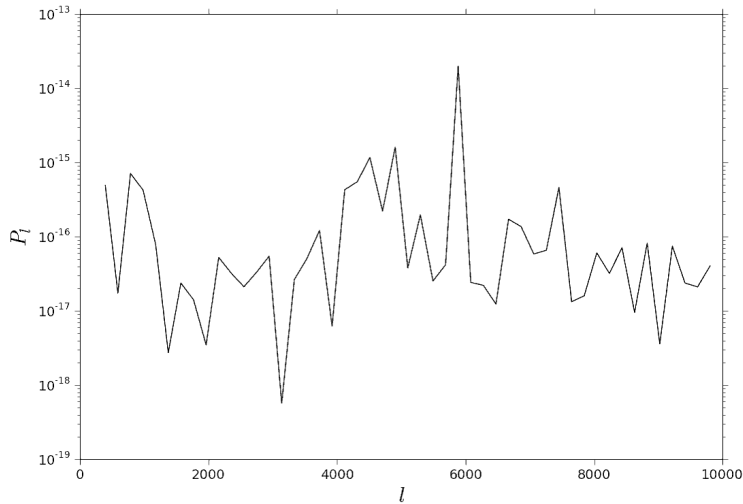}}
\caption{
\label{fig:Pl_PV}
Power spectrum of spurious convergence due to pixel-size variation.
}
\end{figure*}

The bias in cosmological parameters caused by pixel-size variation can be calculated from the spurious convergence as described in Section 3. Figure \ref{fig:Pl_PV} shows spurious convergence power spectrum  $P_l$ used to calculate the bias due to pixel-size variation. The spurious power fluctuates randomly in the vicinity of $10^{-16}$ to $10^{-17}$ and does not show a systematic trend. The convergence power spectrum was calculated on a grid with each element of the grid 30$\times$30 pixels and the convergence averaged over the grid element. This corresponds to 6 arcsec smoothing. 

Table \ref{tab:Bias_PV} shows the amplitude of the bias due to pixel-size variation. 
The  calculated biases shown in Table \ref{tab:Bias_PV}, though larger than those due to tree rings, are very much smaller than the expected marginalized errors for LSST, which as previously stated, are approximately 0.00232 for $\Omega_m$,  0.02434 for $w$ and  0.00427  for $\sigma_8$. 

\begin{table}[htbp]
\centering
\begin{tabular}{|c|c|c|c|}\hline
&$\Omega_m$&$w$&$\sigma_8$\\\hline
bias expectation value		& 1.21E-05 &	-2.18E-05	& -1.79E-05\\\hline
bias standard deviation		& 1.21E-05	& 3.37E-05	& 1.82E-05\\\hline
\textbf{Survey bias }		& 2.42E-05	& 5.55E-05	& 3.60E-05\\\hline
\end{tabular}
\caption{
\label{tab:Bias_PV}
Bias of the cosmological parameters ($\Omega_m, w, \sigma_8$) due to pixel-size variation.
}
\end{table}

\section{Conclusions}

We describe in detail all the steps necessary to go from flat-field images to spurious shear and convergence, then to 2-point correlation functions and spurious power, and finally to bias in cosmological parameters.

For tree rings, we use new formulae developed here to calculate the spurious shear and convergence from the astrometric displacements, and a model developed by \cite{plazas2014}  to estimate the astrometric displacement from the flux modulations measured in flat-field images. 

For the LSST CCDs, the typical value of flux modulation due to tree rings is $0.01\%$, about 50 times smaller than that for the DECam CCDs. The typical amplitude of spurious shear is $0.005\%$ The typical amplitude of the 2-point correlation function over the LSST focal plane is about $10^{-13}$. The 2PCF is much smaller than the square of the typical amplitude of the spurious shear due to cancelations of tangential and radial shears in the calculation. The 2PCF is much smaller than the lensing 2PCF suggesting, but not proving that it can be ignored. We go further and calculate the bias to cosmological parameters and find it to be negligible.

We point out that spurious convergence is the natural way to describe the effect of pixel-size variation, and calculate the spurious convergence, which is directly related to the size of the pixel, from the counts in individual pixels in flat-field images. The width of rows and columns thus determined varies in a periodic way. For the LSST CCDs the typical magnitude of the flux modulation due to periodic pixel-size variation is $0.05\%$, five times larger than for tree rings. This translates into a $10^{-11}$ 2PCF over the LSST focal plane, much smaller than the lensing 2PCF. We calculate the bias to cosmological parameters due to pixel-size variation and find it to be larger than for tree rings, but still negligible compared to the marginalized errors.

In this study we have only considered the impact of tree rings and pixel size variation on cosmological parameters derived from the lensing power spectrum. We plan to extend our analysis to other sensor imperfections.  We note that as much as a factor of 2 additional information about cosmology is contained in  non-Gaussian lensing statistics beyond the power spectrum, we plan to extend our analysis to consider the impact of sensor imperfections on cosmological parameters derived from non-Gaussian lensing statistics.

\acknowledgments{Acknowledgements}

\begin{acknowledgements}
We thank G. Bernstein and P. O'Connor for useful comments and discussions, and the Instrumentation Division of Brookhaven for data taking.
This work was supported in part by the U.S. Department of Energy under Contract No. DE-AC02-98CH10886 and Contract No. DE-SC0012704.
AAP is also supported by JPL, which is run under a contract for NASA by Caltech.
\end{acknowledgements}



\begin{thebibliography}{9}
\bibitem[Chang et al.(2013)]{Chang2013}
Chang C., Kahn S. M., Jernigan J. G., Peterson J. R., AlSayyad Y., Ahmad Z., Bankert J., Bard D., Con- nolly A., Gibson R. R., Gilmore K., Grace E., Hannel M., Hodge M. A., Jee M. J., Jones L., Krughoff S., Lorenz S., Marshall P. J., Marshall S., Meert A., Nagarajan S., Peng E., Rasmussen A. P., Shmakova M., Sylvestre N., Todd N., and Young M., MNRAS 428, 2695 (2013).
\bibitem[Diehl(2012)] {diehl2012}
Diehl H. T., 
Physics Procedia, Proceedings of the 2nd International Conference on Technology and Instrumentation in Particle Physics (TIPP 2011), V37 pp 1332-1340
\bibitem[Dodelson(2003)]{Dodelson2003}
Dodelson Scott,
Modern Cosmology,
Academic Press, 2003
\bibitem[Hamana et al.(2001)]{HamanaMellier}
Hamana T. and Mellier Y.,
Numerical study of the statistical properties of the lensing excursion angles
MNRAS, 2001, 327, 169H
\bibitem[Holland et al.(2014)] {Holland2014}
Holland S. E. , Bebek C. J., Kolbe W. F. and Lee J. S.,
{\emph{Physics of fully depleted CCDs}, (2014), JINST 9 C03057}
\bibitem[Jarvis(2014)]{Jarvis2014}
Jarvis M.,
{\emph{Challenges for precision shape measurements}, (2014), JINST 9 C03017}
\bibitem[Jarvis et al. (2015)]{Jarvis2015} 
Jarvis M., Sheldon E., Zuntz J., Kacprzak T., Bridle S. L., Amara A., Armstrong R., Becker M. R., Bernstein G. M. et al., arXiv:1507.05603(2015).
\bibitem[Kaiser et al.(1993)]{KaiserSquires1993}
Kaiser N., Squires G.
Mapping the dark matter with weak gravitational lensing,
ApJ, 1993, Part 1, vol. 404, no. 2, p. 441-450
\bibitem[Kaiser et al.(2000)]{Kaiser2000}
Kaiser N, Wilson G and Luppino G 2000, arXiv:astro-ph/0003338
\bibitem[Kilbinger et al.(2009)]{Kilbinger2009} 
Kilbinger M. , Benabed K., Guy J., Astier P., Tereno I., Fu L., Wraith D., Coupon J., Mellier Y., Balland C., Bouchet F. R., Hamana T., Hardin D., McCracken H. J., Pain R., Regnault N., Schultheis M., and Yahagi H., Astron. Astrophys.
497, 677 (2009).
\bibitem[Kilbinger(2014)]{Kilbinger2014}
Kilbinger M: arxiv:1411.0115
\bibitem[Lewis et al.(2000)]{CAMB}
Lewis, A., Challinor A. and Lasenby A,
Efficient Computation of {CMB} anisotropies in closed {FRW} models",
ApJ, 2000, 538, 473
\bibitem[Lupton(2014)]{Lupton2014}
Lupton R. ,
{\emph{Consequences of thick CCDs on image processing}, (2014), JINST 9 C04023}
\bibitem[Manfred(1995)]{Manfred1995}
Manfroid J. , 
{\emph{On CCD standard stars and flat-field calibration, (1995)} A\&AS, 113, 587}
\bibitem[Miller et al.(2007)]{Miller2007}
Miller, L., et al., 2007, MNRAS, 382 185
\bibitem[Munshi et al.(2008)]{Munshi2008}
Munshi, D., Valagesas P., Van Waerbeke, L. \& Heavens, A., 2008, Phys. Rept. 462, 67
\bibitem[Petri et al.(2014)]{petri2014}
Petri A., May M., Haiman Z. and Kratochvil J. M.,
Impact of spurious shear on cosmological parameter estimates from weak lensing observables
Physical Review, 2014, D, Volume 90, Issue 12, id.123015 
\bibitem[Plazas et al.(2014)]{plazas2014}
Plazas A. A. et al., Publications of the Astronomical Society of the Pacific, 2014, Volume 126, issue 942, pp.750-760
\bibitem[Rasmussen(2015)] {Rasmussen2015}
Rasmussen A.,
{\emph{Use of sensor characterization data to tune electrostatic model parameters for LSST sensors}, (2015), JINST 10 C05028}
\bibitem[Schneider(2006)]{Schneider2006}
Schneider, 2006, Part 1: Introduction to gravitational lensing and cosmology, ISBN: 3-540-30309-X
\bibitem[Smith et al.(2008)] {Smith2008}
Smith R. and Rahmer G.,
{\emph{Pixel area variation in CCDs and implications for precision photometry}, (2008), Proc. SPIE 7021, 70212A-1}
\bibitem[Springel(2005)]{Gadget-2}
Springel V.,
The cosmological simulation code GADGET-2,
MNRAS, 2005, 364, 1105S
\bibitem[Stubbs(2014)]{Stubbs2014}
Stubbs C. ,
{\emph{Precision astronomy with imperfect fully depleted CCDs: an introduction and a suggested lexicon, (2014), JINST 9 C03032} arXiv:1312.2313}
\bibitem[Tucker et al.(2007)]{Tucker2007}
Tucker D. et al., 
The Future of Photometric, Spectrophotometric and Polarimetric Standardization, ASP Conference Series, Vol. 364, Proceedings of a conference held 8-11 May, 2006 in Blankenberge, Belgium. Edited by C. Sterken. San Francisco: Astronomical Society of the Pacific, 2007., p.187
\bibitem[Wittman et al.(2000)]{Wittman2000}
Wittman D. M., Tyson J. A., Kirkman D., Dell'Antonio, I. and Bernstein, G. 2000, Nature, 405, 143-148.
\end{thebibliography}
\end{document}